\documentclass[aps,prl,twocolumn,superscriptaddress,preprintnumbers]{revtex4-1}
\usepackage{amssymb}
\usepackage{graphicx}
\usepackage{amsmath}
\usepackage{hyperref}
\usepackage{subfigure}
\usepackage{multirow}
\usepackage{setspace}
\usepackage{verbatim}
\usepackage{float}
\usepackage{color}
\usepackage{ulem}
\usepackage[utf8]{inputenc}
\usepackage[table,xcdraw]{xcolor}
\usepackage{makecell}
\usepackage{booktabs}
\usepackage{graphicx}
\usepackage{dcolumn}
\usepackage{bm}
\usepackage{mathrsfs}
\usepackage{xcolor,graphicx}
\usepackage{dcolumn}
\usepackage{bm}
\usepackage{enumerate}
\usepackage{slashed}
\usepackage{subfigure}
\usepackage{todonotes}
\usepackage{multirow}
\usepackage{hyperref}
\usepackage{xspace}
\usepackage{float}
\usepackage{ulem}
\usepackage[T1]{fontenc}

\begin{document}
	
	\title{First Constraint on Axion-Photon Coupling $g_{\gamma}$ from Neutron Star Observations}
	
	\author{Jun-Chen Wang}
	\email{junchenwang@stu.pku.edu.cn}
	\affiliation{School of Physics, Peking University, Beijing 100871, China}

    \author{Shunshun Cao}
    \email{19css@pku.edu.cn}
\affiliation{School of Physics, Peking University, Beijing 100871, China}
\affiliation{State Key Laboratory of Nuclear Physics and Technology, School of Physics, Peking University}

    \author{Jinchen Jiang}
    \email{jjiang@mpifr-bonn.mpg.de}
    \affiliation{Max-Planck institut f\"ur Radioastronomie, Auf Dem H\"ugel, Bonn, 53121, Germany}
     \affiliation{National Astronomical Observatories, Chinese Academy of Sciences, Beijing 100012, China}

    \author{Yandong Liu}
    \email{ydliu@bnu.edu.cn (corresponding author)}
    \affiliation{Key Laboratory of Beam Technology of Ministry of Education, School of Physics and Astronomy, Beijing Normal University, Beijing, 100875, China}
    \affiliation{Institute of Radiation Technology, Beijing Academy of Science and Technology, Beijing 100875, China}

    \author{Qing-Hong Cao}
    \email{qinghongcao@pku.edu.cn (corresponding author)}
    \affiliation{School of Physics, Peking University, Beijing 100871, China}
    \affiliation{Center for High Energy Physics, Peking University, Beijing 100871, China}

    \author{Lijing Shao}
    \email{lshao@pku.edu.cn (corresponding author)}
    \affiliation{Kavli Institute for Astronomy and Astrophysics, Peking University, Beijing 100871, China}
    \affiliation{National Astronomical Observatories, Chinese Academy of Sciences, Beijing 100012, China}

\hyphenpenalty=3000
\hbadness=3000

\preprint{CPTNP-2025-017}

    \begin{abstract}
We propose a novel method to detect axions which uniquely depends on the dimensionless axion-photon coupling $g_{\gamma}$, independent of the suppressive axion decay constant $f_a$.
Using neutron star PSR B1919+21 data from the Five-hundred-meter Aperture Spherical Telescope, we derive the first constraint $|g_{\gamma}|<0.93$ at $1\sigma$ confidence level for ultra-light axions ($m_a < 10^{-11}$ eV). 	
\end{abstract}
\maketitle

\noindent\textit{\textbf{Introduction}}--The axion is a well-motivated hypothetical particle originally proposed to resolve the strong CP problem~\cite{Crewther:1979pi,Pendlebury:2015lrz,Peccei:1977hh,Peccei:1977ur,Weinberg:1977ma,Wilczek:1977pj,Kim:1979if,Shifman:1979if,Dine:1981rt,Kim:2008hd}. Its low-energy interactions are described by the effective Lagrangian
\begin{align}
\label{eq:axionint}
\mathcal{L} = \frac{\alpha_s}{8\pi} \frac{a}{f_a} G^b_{\mu\nu} \tilde{G}^{b\mu\nu} + \frac{1}{4} \frac{g_\gamma}{f_a} a F^{\mu\nu} \tilde{F}_{\mu\nu},
\end{align}
where $a$ is the axion field, $f_a$ the decay constant, and $g_\gamma$ a dimensionless, quantized axion-photon coupling of particular theoretical interest~\cite{Srednicki:1985xd}. Recent studies have linked $g_\gamma$ to global aspects of the Standard Model (SM) gauge group and generalized symmetries~\cite{Hucks:1990nw,Tong:2017oea,Brennan:2023mmt,Choi:2023pdp}.

Most current axion searches focus on small fluctuations around the vacuum expectation value (VEV), leading to signals proportional to $g_\gamma/f_a$ and thus suppressed by large $f_a$. These include storage-ring~\cite{Graham:2020kai,Agrawal:2022wjm,Dror:2022xpi,JEDI:2022hxa}, cavity-based~\cite{ADMX:2009iij,Brubaker:2016ktl,Geraci:2018fax}, axion-photon mixing experiments~\cite{Raffelt:1987im,Galanti:2022yxn,Cao:2023kdu} and cosmological observations~\cite{Li:2024bbe}. Alternatively, scenarios exploiting distinct axion VEVs, such as cosmological birefringence~\cite{Hoseini:2019woh,Fujita:2020ecn,Agrawal:2019lkr}, domain wall networks~\cite{Sikivie:1982qv,Pospelov:2012mt,Kawasaki:2014sqa}, and Berry-phase measurements~\cite{Cao:2024lwg}, enable probes of quantized $g_{\gamma}$ independently of $f_a$, though they often require complex cosmic topological structures~\cite{Reece:2023iqn,Agrawal:2023sbp,Cordova:2023her}.
\begin{figure}[h]
    \centering
    \includegraphics[width=1.0\linewidth]{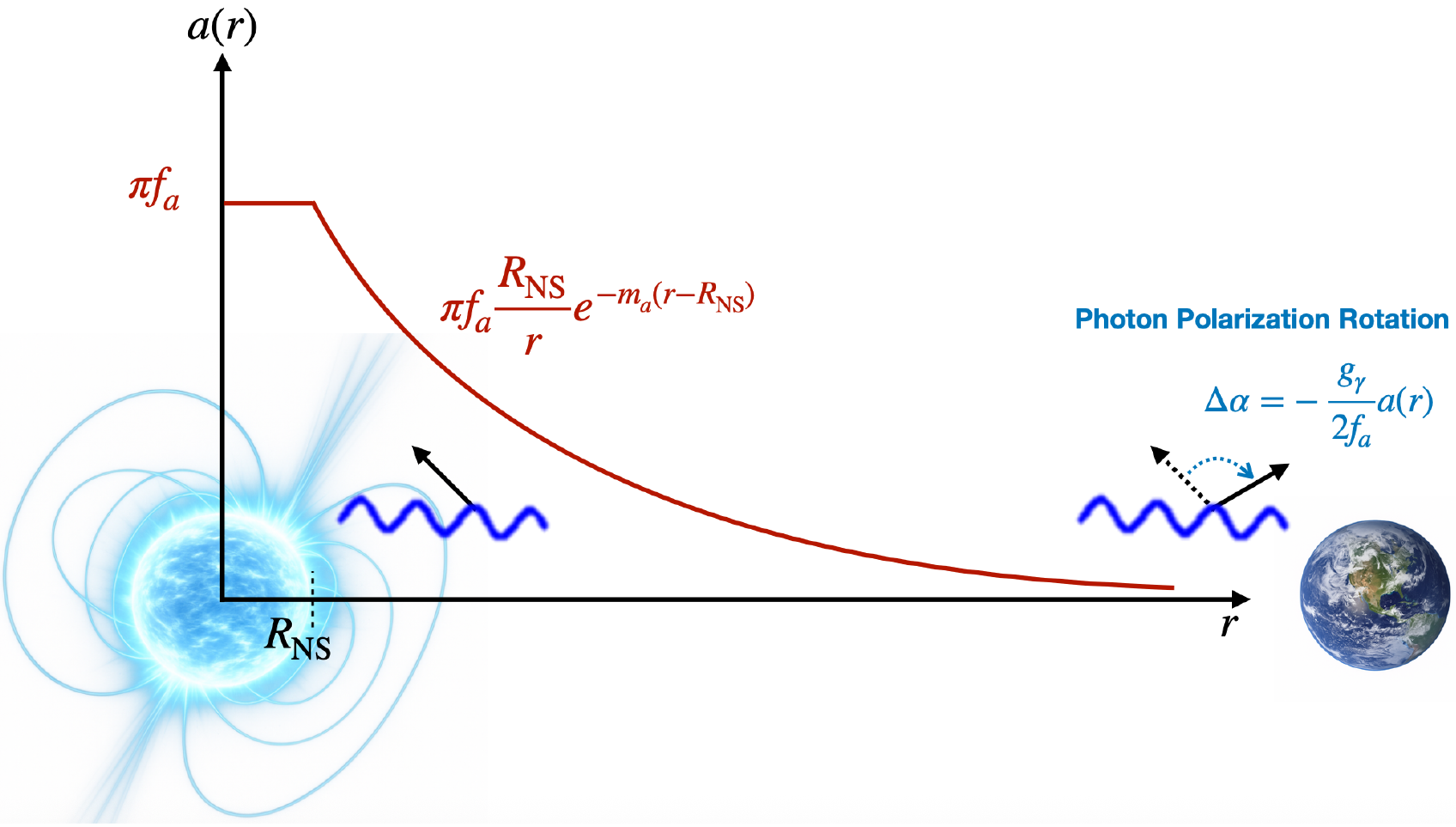}
    \caption{Illustration of the axion field profile $a(r)$ induced by chiral symmetry restoration inside a neutron star, and the resulting polarization rotation \( \Delta \alpha \) experienced by radio photons as they propagate through the axion background toward Earth. }
    \label{fig:s-axion}
\end{figure}

In this Letter, we propose a novel observational strategy based on neutron stars, where high nucleon densities restore QCD chiral symmetry, effectively nullifying the $G\tilde{G}$ term in Eq.~\eqref{eq:axionint}. As a result, the axion field undergoes a phase transition to a new VEV. This transition induces a macroscopic axion configuration sourced by the neutron star; see Fig.~\ref{fig:s-axion}. Employing Radius-Frequency Mapping (RFM)~\cite{Ruderman:1975ju,lesch1994radiative,qiu2023rotating}, we identify a correlation between axion-induced modifications of photon polarization angles and radio emission frequencies, enabling a direct probe of $g_\gamma$.

The resulting axion field profile is given by~\cite{Hook:2017psm,Huang:2018pbu,Kumamoto:2024wjd},
\begin{align}
\label{eq:profile2}
a(r) = \begin{cases}
\pi f_a, & r < R_{\rm NS}, \\
\pi f_a \frac{R_{\rm NS}}{r} e^{-m_a(r - R_{\rm NS})}, & r > R_{\rm NS},
\end{cases}
\end{align}
where $m_a$ is the axion mass and $R_{\rm NS}$ the neutron star radius; see Supplemental Material for details. This Yukawa-like profile outside the star effectively endows it with an “axion charge” $\sim \pi f_a R_{\rm NS}$, offering a new experimental handle on axion physics. This mechanism provides a unique observational window into ultraviolet axion completions~\cite{Zhitnitsky:1980tq,Dine:1981rt,DiLuzio:2020wdo,Kim:1979if,Shifman:1979if,Srednicki:1985xd,Agrawal:2022lsp,DiLuzio:2017pfr,DiLuzio:2016sbl}, the global structure of the SM~\cite{Hucks:1990nw,Tong:2017oea,Brennan:2023mmt}, and associated generalized symmetries~\cite{Choi:2023pdp}—all without reliance on exotic cosmological relics or ultra-high precision laboratory setups.

\noindent \textit{\textbf{Axion detection}}--The axion-photon interaction in Eq.~\eqref{eq:axionint} induces photon birefringence, extensively studied previously~\cite{Hoseini:2019woh,Fujita:2020ecn,Carroll:1989vb,Harari:1992ea,Agrawal:2019lkr,Fedderke:2019ajk,Sikivie:2020zpn}. Linearly polarized photons traveling through an axion background experience polarization rotation by~\cite{Hoseini:2019woh,Fujita:2020ecn,Carroll:1989vb,Harari:1992ea,Agrawal:2019lkr,Fedderke:2019ajk,Sikivie:2020zpn}
\begin{equation}
\label{eq:rotation_angle}
\Delta \alpha = \frac{g_{\gamma}(a_f - a_i)}{2f_a},
\end{equation}
where $a_i$ and $a_f$ denote axion field values at photon emission and detection points. Setting $a_f=0$ at Earth and using the neutron star axion profile in Eq.~\eqref{eq:profile2}, the rotation angle explicitly depends on radial distance $r$ from the neutron star:
\begin{equation}
\label{eq:rotation_angle_final}
\Delta \alpha(r) = -\frac{g_{\gamma}}{2f_a}a(r).
\end{equation}

To circumvent uncertainties in photon emission radius measurements, we apply RFM to relate photon emission radius directly to photon frequency $\omega_c$ (derived in Appendix \ref{app:RFM}):
\begin{equation}
\label{eq:omega_value}
\omega_c=\frac{9}{8}(2\pi)^{\frac{1}{2}}\gamma^3P_{\rm NS}^{-\frac{1}{2}}r^{-\frac{1}{2}}\simeq 2.82\frac{ \gamma^3}{\sqrt{P_{\rm NS}r}},
\end{equation}
where $P_{\rm NS}$ is the neutron star rotation period, and $\gamma$ is the Lorentz factor of the magnetospheric plasma~\cite{Ruderman:1975ju}. Given the frequency-independent nature of initial polarization angles~\cite{2012puas.bookL}, we introduce a constant initial angle $\alpha_0$, fitting this parameter from observational data. The observed polarization angle $\alpha_{\rm obs}$ relates via $\Delta \alpha=\alpha_{\rm obs}-\alpha_0$.

Combining Eqs.~\eqref{eq:profile2}, \eqref{eq:rotation_angle_final}, and \eqref{eq:omega_value}, we obtain a direct relation between polarization rotation $\Delta \alpha$ and photon frequency $\omega_c$:
\begin{equation}
\label{eq:relationship_final}
\Delta \alpha(\omega_{c})=-\frac{\pi g_{\gamma}}{2}\frac{R_{\rm NS}P_{\rm NS}\omega_{c}^2}{7.95 \gamma^6}e^{\left[-m_a\left(\frac{7.95\gamma^6}{P_{\rm NS}\omega_c^2}-R_{\rm NS}\right)\right]}.
\end{equation} 
This result reveals a key advantage of the proposed method: 
the polarization rotation $\Delta \alpha$ depends solely on the dimensionless axion-photon coupling $g_{\gamma}$, independently of the axion decay constant $f_a$. This independence arises because the axion vacuum expectation value (VEV), given by Eq.~\eqref{eq:profile2}, scales linearly with $f_a$, thereby canceling the $f_a$ dependence in the rotation angle.

\begin{figure}
		\includegraphics[width=0.49\textwidth]{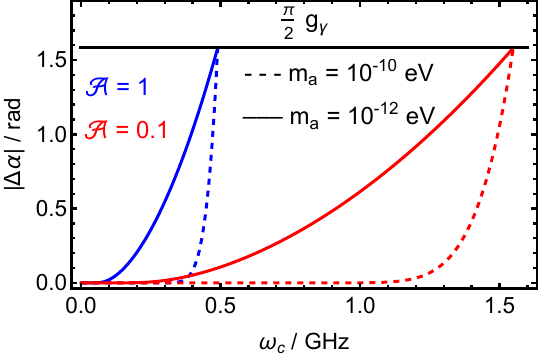}
		\caption{Polarization rotation angle $\Delta \alpha$ as a function of photon frequency $\omega_c$, illustrating the effect of neutron star and axion parameters. The dimensionless coupling is set to $g_{\gamma} = 1$, with neutron star radius $R_{\rm NS} = 10~\mathrm{km}$ and particle Lorentz factor $\gamma = 100$. Curves are shown for spin periods  $P_{\rm NS} = 1~\mathrm{s}$ (blue) and $0.1~\mathrm{s}$ (red), and axion masses $m_a = 10^{-10}~\mathrm{eV}$ (solid) and $10^{-12}~\mathrm{eV}$ (dashed).}
		\label{fig:profile}
	\end{figure}

For practical measurements, Eq.~\eqref{eq:relationship_final} simplifies to
\begin{equation}
\label{eq:constraint_full}
		\Delta \alpha =-6.59g_{\gamma}\mathcal{A}\left(\frac{\omega_c}{\mathrm{GHz}}\right)^2e^{-m_a R_{\rm NS}\left[\frac{0.24}{\mathcal{A}}\left(\frac{\mathrm{GHz}}{\omega_c}\right)^2-1\right]}
\end{equation}
with
\begin{equation}
\label{eq:A_define}
\mathcal{A}=\left(\frac{R_{\rm NS}}{10~\mathrm{km}}\right)\left(\frac{P_{\rm NS}}{1~\mathrm{s}}\right)\left(\frac{100}{\gamma}\right)^6.
\end{equation}
Figure~\ref{fig:profile} illustrates the frequency-dependent polarization rotation, indicating clear upper limits for both $\omega_c$ and $\Delta \alpha$. The maximum photon frequency at the neutron star surface is
\begin{equation}
\label{eq:omega_max}
\omega_{\rm max}=\sqrt{\frac{0.24}{\mathcal{A}}}~\mathrm{GHz},
\end{equation}
with the maximum rotation angle
\begin{equation}
\label{eq:alpha_max}
|\Delta \alpha|_{\rm max}=\frac{\pi}{2}g_{\gamma},
\end{equation}
independent of $\mathcal{A}$ and $m_a$ (black line in Fig.~\ref{fig:profile}). For ultralight axions ($m_a<10^{-11} \mathrm{eV}$), the exponential factor becomes negligible, yielding
\begin{equation}
\label{eq:constraint_light}
\Delta \alpha=-6.59 g_{\gamma}\mathcal{A}\left(\frac{\omega_c}{\mathrm{GHz}}\right)^2,
\end{equation}
shown as colored solid lines in Fig.~\ref{fig:profile}. In summary, measurements of polarization rotation directly constrain the coupling $g_{\gamma}$ and, through frequency-dependent fitting, provide sensitivity to axion masses $m_a$.

\noindent \textit{\textbf{Constraints from FAST data}} -- Using the theoretical axion-induced polarization-frequency relationship from Eq.~\eqref{eq:constraint_full}, we constrain axion parameters $g_{\gamma}$ and $m_a$ using observational data from pulsar PSR B1919+21 (J1921+2153), recorded by the Five-hundred-meter Aperture Spherical radio Telescope (FAST) on 2023 August 23~\footnote{The raw data are obtained from the FAST data center
(https://fast.bao.ac.cn/cms/category/data$\_$center$\_$en).}. Observations utilized the central beam of FAST’s L-band 19-beam receiver over a 1200 s duration~\cite{2019SCPMA..6259502J,Jiang_2020,Melo:2020ogc}. Both orthogonal linear polarizations were recorded, covering a 500 MHz bandwidth (1–1.5 GHz), channelized into 4096 frequency channels, each sampled every $49.152~\mu\mathrm{s}$.

 Polarization calibration employed one-minute noise injections at observation start and end, correcting differential gain and delays between linear polarizations via a single-axis model. Data processing utilized software packages \texttt{DSPSR}~\cite{vanStraten:2010hy}, \texttt{PSRCHIVE}~\cite{Hotan:2004tz}, and \texttt{TEMPO2}~\cite{Hobbs:2006cd}. Faraday rotation ($\Delta\alpha=\mathrm{RM}\times\lambda^2$) was characterized with rotation measure $\mathrm{RM}=-13.1 \pm 1.1\,\mathrm{rad\,m^{-2}}$, implying roughly a $3^\circ$ polarization angle uncertainty across the observed band~\cite{Cao:2024sdx}. The processed dynamic spectra provided normalized Stokes parameters, $Q=\cos(\alpha_{\rm obs})$ and $U=\sin(\alpha_{\rm obs})$, with uncertainties estimated from the white noise in profile baseline. The possible systematic errors caused by RM uncertainty is not included.

\begin{figure}
		\includegraphics[width=0.49\textwidth]{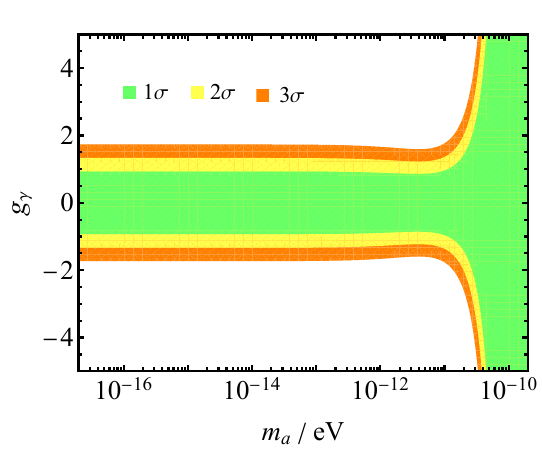}
		\caption{Constraints on the axion-photon coupling $g_{\gamma}$ and axion mass $m_a$ from radio polarization data of the neutron star PSR B1919+21 (J1921+2153), based on a $\chi^2$ fit to FAST data. The green, yellow, and orange contours denote the $1\sigma$, $2\sigma$ and $3\sigma$ confidence level regions, respectively.}
		\label{fig:parameter}
\end{figure}

Axion parameters were constrained via $\chi^2$ fitting, involving four free parameters: $g_{\gamma}$, $m_a$, pulsar factor $\mathcal{A}$, and initial polarization angle $\alpha_0$. The theoretical Stokes parameters are:
\begin{align}
        \label{eq:Q_th}
        &Q_{\rm th}(\omega_c, g_{\gamma}, m_a, \mathcal{A}, \alpha_0) = \\ 
        &\notag \cos \left(-6.59g_{\gamma}\mathcal{A}\left(\frac{\omega_c}{\mathrm{GHz}}\right)^2e^{-m_a R_{\rm NS}\left[\frac{0.24}{\mathcal{A}}\left(\frac{\mathrm{GHz}}{\omega_c}\right)^2-1\right]}+\alpha_0\right), \\
        \label{eq:U_th}
        & U_{\rm th}(\omega_c, g_{\gamma}, m_a, \mathcal{A}, \alpha_0) = \\
        \notag 
        &\sin \left(-6.59g_{\gamma}\mathcal{A}\left(\frac{\omega_c}{\mathrm{GHz}}\right)^2e^{-m_a R_{\rm NS}\left[\frac{0.24}{\mathcal{A}}\left(\frac{\mathrm{GHz}}{\omega_c}\right)^2-1\right]}+\alpha_0\right),
    \end{align}
where $R_{\rm NS}=10$ km. The fitting statistic is
  \begin{equation}
        \label{eq:chi_2_define}
        \chi^2 = \sum_{i} \left(\frac{Q_{\rm data}^{i}-Q_{\rm th}^{i}}{Q_{\rm er}^{i}} \right)^2+ \sum_{i} \left(\frac{U_{\rm data}^{i}-U_{\rm th}^{i}}{U_{\rm er}^{i}} \right)^2,
    \end{equation}
where $i$ is the index of frequency channel. Minimizing $\chi^2$, we derive constraints illustrated in Fig.~\ref{fig:parameter}. For axions with $m_a<10^{-11}$ eV, the coupling $g_{\gamma}$ is effectively constrained due to negligible exponential suppression, yielding mass-independent limits at $1\sigma$, $2\sigma$, and $3\sigma$ confidence levels of $|g_{\gamma}|<0.93$, 1.33, and 1.73 respectively. For $m_a>10^{-10}$ eV, sensitivity declines significantly due to exponential suppression.

    \noindent \textit{\textbf{Conclusion}}--We have demonstrated that the axion phase transition induced by neutron stars provides a novel observational approach for detecting axions. Employing photon birefringence caused by axion fields and pulsar RFM, we predict observable, frequency-dependent polarization rotations in neutron star radio emissions. These measurements directly constrain the dimensionless axion-photon coupling $g_{\gamma}$, independent of the traditionally suppressive axion decay constant $f_a$, thus enhancing detection sensitivity. 
 Our analysis of FAST data constrains the axion-photon coupling to $|g_{\gamma}|<0.93$, 1.33 and 1.93 at $1\sigma$, $2\sigma$ and $3\sigma$ confidence levels for axion masses below $10^{-11}$ eV, respectively.

\noindent 
\textbf{Acknowledgments}--This work made use of the data from FAST (Five-hundred-meter Aperture Spherical radio Telescope)(https://cstr.cn/31116.02.FAST).  
FAST is a Chinese national mega-science facility, operated by National Astronomical Observatories, Chinese Academy of Sciences. The work of J.W., Y.L. and Q.C. is partly supported by the National Science Foundation of China under Grant Nos. 12075257, 12175016 and 12235001.
The work of J.J. is partly supported by the National SKA Program of China (2020SKA0120100) and the European Research Council (ERC) starting grant ``COMPACT" (101078094).
The work of L.S. is partly supported by the National SKA Program of China (2020SKA0120300), the Beijing Natural Science Foundation (1242018), and the Max Planck Partner Group Program funded by the Max Planck Society. The authors gratefully acknowledge the valuable discussions and insights provided by the members of the Collaboration on Precision Tests and New Physics (CPTNP).

\bibliographystyle{apsrev4-1}
\bibliography{ref}

\begin{appendix}

\section{Axion phase transition in neutron stars and corresponding profile}
The axion profile presented in the main text plays a crucial role in our analysis. While previous works~\cite{Hook:2017psm,Huang:2018pbu,Kumamoto:2024wjd} derived this profile by assuming a suppressed axion mass, we provide here a more general derivation starting from the axion potential. We begin with the QCD-induced axion potential~\cite{GrillidiCortona:2015jxo}:
\begin{equation}
V_0 = -m_{\pi}^2 f_{\pi}^2 \sqrt{1 - \frac{4 m_u m_d}{(m_u + m_d)^2} \sin^2\left(\frac{a}{2f_a}\right)},
\end{equation}
where \( m_{\pi} \) and \( f_{\pi} \) are the pion mass and decay constant, and \( m_u \), \( m_d \) are the up and down quark masses. For simplicity, we take \( m_u = m_d \equiv m_q \). Using chiral perturbation theory, \( m_\pi^2 = -2 m_q \langle\bar{q} q\rangle / f_\pi^2 \), with \( \langle\bar{q} q\rangle \equiv \langle\bar{u} u + \bar{d} d\rangle / 2 < 0 \). The potential becomes
\begin{equation}
V_0 = 2 m_q \langle\bar{q} q\rangle \left|\cos\left(\frac{a}{2f_a}\right)\right|.
\end{equation}
This has a minimum at \( a = 0 \), consistent with bounds from EDM measurements~\cite{Crewther:1979pi,Pendlebury:2015lrz}.

However, in many UV-complete models, higher-dimensional operators can significantly modify the axion potential~\cite{Banks:2010zn,Witten:2017hdv,Harlow:2018tng,Alvey:2020nyh,Dine:2022mjw}. Within effective field theory, these effects are captured by~\cite{Alvey:2020nyh,Dine:2022mjw}:
\begin{equation}
\Delta V = \Lambda^4 \sum_{n=1}^\infty 2^{1 - n/2} |\lambda_n| \left(\frac{f_a}{\Lambda}\right)^n \cos\left(n \frac{a}{f_a} + \beta_n\right),
\end{equation}
where \( \Lambda > f_a \) is the cutoff scale and \( |\lambda_n| \), \( \beta_n \) are the magnitude and phase of the effective coupling. Since \( f_a / \Lambda < 1 \), the \( n = 1 \) term dominates. The full potential is thus approximated as
\begin{align}
V(a) &\simeq 2 m_q \langle\bar{q} q\rangle \left|\cos\left(\frac{a}{2f_a}\right)\right| \notag \\
&\quad + \sqrt{2} \Lambda^3 f_a |\lambda_1| \cos\left(\frac{a}{f_a} + \beta_1\right).
\end{align}
Requiring that the global minimum remains at \( a = 0 \), consistent with experiments~\cite{Crewther:1979pi,Pendlebury:2015lrz}, yields two scenarios:
\begin{align}
\text{(I)}\quad &\beta_1 = \pi, \\
\text{(II)}\quad &\beta_1 = 0,\quad m_q |\langle\bar{q} q\rangle| > 2\sqrt{2} \Lambda^3 f_a |\lambda_1|.
\end{align}
We focus on Scenario II, which allows for axion phase transitions in dense environments.

Such a transition arises when QCD chiral symmetry, spontaneously broken in vacuum, is restored at high nucleon densities~\cite{Pietroni:2005pv,Olive:2007aj,Hinterbichler:2010es}. Let \( \langle\bar{q} q\rangle_{\rho_N} \) denote the in-medium condensate at nucleon density \( \rho_N \). Studies indicate \( |\langle\bar{q} q\rangle_{\rho_N}| < |\langle\bar{q} q\rangle_0| \)~\cite{Cohen:1991nk,Brown:1991kk}. Thus, if
\begin{equation}
m_q |\langle\bar{q} q\rangle_{\rho_N}| < 2\sqrt{2} \Lambda^3 f_a |\lambda_1| < m_q |\langle\bar{q} q\rangle_0|,
\end{equation}
then the axion vacuum shifts from \( a = 0 \) (in vacuum) to
\begin{equation}
a_{\rho_N} = 2f_a \arccos\left( \frac{m_q |\langle\bar{q} q\rangle_{\rho_N}|}{2\sqrt{2} \Lambda^3 f_a |\lambda_1|} \right).
\end{equation}
In neutron stars, where \( \rho_N^{\text{NS}} \sim 10^6\text{--}10^9\ \text{MeV}^3 \)~\cite{Jamin:2002ev,Moodley:2024ech}, chiral symmetry is restored and \( a_{\rm NS} = \pi f_a \). On Earth, \( \rho_N^{\text{Earth}} \sim 3 \times 10^{-8}\ \mathrm{MeV}^3 \), and \( \langle\bar{q} q\rangle \approx \langle\bar{q} q\rangle_0 \).

We now derive the axion field profile around the neutron star. The equation of motion is
\begin{align}
\partial_\mu \partial^\mu a = -\frac{dV}{da},
\end{align}
with \( V(a) \) as above. Inside the star, where \( \langle\bar{q} q\rangle = 0 \), the solution \( a = \pi f_a \) satisfies the equation. Outside, we expand the potential:
$V(a)\sim 2m_q \left<\overline{q}q\right>_0+\sqrt{2}\Lambda^3f_a|\lambda_1|+m_a^2a^2/2$ with $m_a^2 \equiv -m_q\left<\overline{q}q\right>_0/(2f_a^2)-\sqrt{2}\Lambda^3|\lambda_1|/f_a$.
The radial equation becomes
\begin{equation}
\frac{1}{r^2} \frac{d}{dr} \left(r^2 \frac{da}{dr} \right) = m_a^2 a,
\end{equation}
with Yukawa-like solution
\begin{equation}
a(r) = C \frac{e^{-m_a r}}{r}. \nonumber
\end{equation}
Matching at \( r = R_{\rm NS} \) gives \( C = \pi f_a R_{\rm NS} e^{m_a R_{\rm NS}} \). The full profile is
\begin{equation}
a(r) = \begin{cases}
\pi f_a, & r < R_{\rm NS}, \\
\pi f_a \dfrac{R_{\rm NS}}{r} e^{-m_a (r - R_{\rm NS})}, & r > R_{\rm NS}.
\end{cases}
\end{equation}
This profile has a Yukawa form, with effective ``axion charge" \( \pi f_a R_{\rm NS} \)~\cite{Hook:2017psm,Huang:2018pbu}.

\section{Derivation of the Radius-Frequency Mapping}\label{app:RFM}
In this section, we derive the expression for the Radius-Frequency Mapping (RFM) presented in the main text. The key physics underlying RFM originates from the photon emission mechanism in neutron star magnetospheres. Neutron star radio emission is primarily attributed to curvature radiation from relativistic electrons, positrons, and ions. The characteristic frequency of curvature radiation is given by~\cite{Ruderman:1975ju,lesch1994radiative,qiu2023rotating}
\begin{equation}
\label{eq:characteristic_frequency}
\omega_c = \frac{3\gamma^3}{2\rho},
\end{equation}
where \( \rho \) is the curvature radius of the particle trajectory, and \( \gamma \) is the Lorentz factor. Typical values are $\gamma \sim 10^2\text{--}10^3$~\cite{Ruderman:1975ju,lesch1994radiative,qiu2023rotating}.

To relate \( \omega_c \) to the radial distance \( r \), we invoke two widely accepted assumptions. First, the magnetic field of the neutron star is approximated as a dipole~\cite{radhakrishnan1969magnetic,Ruderman:1975ju,Xue:2023ejt}. Second, radiation originates from particles moving along the last open magnetic field lines~\cite{Ruderman:1975ju,qiu2023rotating}. Under these conditions, the curvature radius is~\cite{qiu2023rotating}
\begin{equation}
\label{eq:rho_value}
\rho = \frac{4}{3} \sqrt{r_e r},
\end{equation}
where \( r_e \) denotes the maximum radial extension of the last open field line. This is approximately given by
\begin{equation}
\label{eq:re_value}
r_e \simeq \frac{P_{\rm NS}}{2\pi},
\end{equation}
with \( P_{\rm NS} \) the neutron star rotation period. Substituting Eqs.~\eqref{eq:rho_value} and~\eqref{eq:re_value} into Eq.~\eqref{eq:characteristic_frequency}, we obtain the mapping between frequency and emission radius:
\begin{equation}
\label{eq:omega_value_app}
\omega_c = \frac{9}{8}(2\pi)^{1/2} \gamma^3 P_{\rm NS}^{-1/2} r^{-1/2} \simeq 2.82 \frac{\gamma^3}{\sqrt{P_{\rm NS} r}}.
\end{equation}
This relation encodes the key idea of RFM: higher-frequency radiation originates from lower altitudes in the neutron star magnetosphere.

\end{appendix}

\end{document}